\documentclass[11pt]{article}
\usepackage{amsmath}
\usepackage{amsthm}
\usepackage{amssymb}
\usepackage{graphicx}
\usepackage{url}
\usepackage{bm}
\usepackage[comma,authoryear]{natbib}

\usepackage{graphicx,type1cm,eso-pic,color}

\numberwithin{equation}{section}

\begin{document}
\setcounter{page}{1}
\thispagestyle{empty}
\markboth{}{}
\newcommand{\twopartdef}[4]
{
	\left\{
		\begin{array}{ll}
			#1 & \mbox{} #2 \\
			#3 & \mbox{} #4
		\end{array}
	\right.
} 

\pagestyle{myheadings}
\markboth{Contamination mapping using a multivariate spatial Bayesian model}{Contamination mapping using a multivariate spatial Bayesian model}

\date{}




\baselineskip 20truept

\begin{center}
{\Large {\bf Contamination mapping in Bangladesh using a multivariate spatial Bayesian model for left-censored data}} 
\end{center}

\vspace{.1in} 

\begin{center}
{\large {\bf Indranil Sahoo}} \\
{\large {\it Department of Statistical Sciences and Operations Research,\\
Virginia Commonwealth University, Richmond, United States}} \\
{\large {\bf Arnab Hazra}}\\
{\large {\it Computer, Electrical and Mathematical Sciences and Engineering Division, \\King Abdullah University of Science and Technology, Thuwal, Saudi Arabia.}}
\end{center}

\vspace{.1in}
\baselineskip 18truept

\begin{abstract}

Arsenic (As) and other toxic elements contamination of groundwater in Bangladesh poses a major threat to millions of people on a daily basis. {\color{black} Understanding complex relationships between arsenic and other elements can provide useful insights for mitigating arsenic poisoning in drinking water and it requires multivariate modeling of the elements.} However, environmental monitoring of such contaminants often involves a substantial proportion of left-censored observations falling below a minimum detection limit (MDL). This problem motivates us to propose a multivariate spatial Bayesian model for left-censored data for investigating the abundance of arsenic in Bangladesh groundwater and for creating spatial maps of the contaminants. Inference about the model parameters is drawn using an adaptive Markov Chain Monte Carlo (MCMC) sampling. The computation time for the proposed model is of the same order as a multivariate Gaussian process model that does not impute the censored values. The proposed method is applied to the arsenic contamination dataset made available by the Bangladesh Water Development Board (BWDB). Spatial maps of arsenic, barium (Ba), and calcium (Ca) concentrations in groundwater are prepared using the posterior predictive means calculated on a fine lattice over Bangladesh. Our results indicate that Chittagong and Dhaka divisions suffer from excessive concentrations of arsenic and only the divisions of Rajshahi and Rangpur have safe drinking water based on recommendations by the World Health Organization (WHO).

\end{abstract}

\vspace{.1in} 



\section{Introduction}
\label{introduction}

Arsenic contamination in groundwater is a type of water pollution that is often due to naturally occurring high concentrations of arsenic in the soil. The presence of an abundant quantity of arsenic in groundwater is now a common problem in various parts of the world, including Argentina, Bangladesh, Chile, China, Hungary, India, Mexico, Nepal, Taiwan, and the USA \citep{hossain2006arsenic, bagchi2007arsenic}. However, the contamination of groundwater by naturally occurring inorganic arsenic in Bangladesh is reported as the largest environmental arsenic poisoning of a population in history \citep{smith2000contamination, bagchi2007arsenic}. It is a high-profile problem due to the abundant use of deep tube wells for water supply in the Ganges Delta. The scale of this environmental poisoning disaster is said to be greater than the accident in Bhopal, India in 1984, and Chernobyl, Ukraine, in 1986 \citep{pearce2001bangladesh}. The first case of arsenic poisoning was identified by the Department of Public Health Engineering (DPHE), Bangladesh in 1993 \citep{chakraborti2015groundwater}. Currently, the situation in Bangladesh is dire, with at least 50 of the 64 districts reportedly suffering from arsenic contamination, and an estimated 50 million inhabitants are at risk of drinking contaminated water \citep{ahamed2006eight, ravenscroft2011arsenic}. Over the past two decades, there has been a plethora of research on arsenic contamination in Bangladesh, including studying the extension of contamination, numerous health consequences, and possible mitigation strategies. See \cite{yunus2016review} for a complete review of research in this regard. As mentioned in \cite{yunus2016review}, research efforts regarding arsenic contamination in Bangladesh have diminished over the years but the issue still persists.


Over the past two decades, several geostatistical models have been used to predict arsenic concentration at unobserved locations in different countries \citep{goovaerts2005geostatistical, lee2007evaluation, jangle2016statistical}. Spatial distribution and spatial variability of arsenic concentration in the groundwater of Bangladesh have also been studied in \cite{karthik2001spatial, serre2003application, gaus2003geostatistical, hossain2007geostatistically, winkel2008predicting}. In most scenarios, instruments used to measure arsenic and other contaminants have detection limits. The data falling below (above) some lower (upper) detection limits are censored, and the exact measurements are not available. Usually, arsenic concentrations in groundwater that fall below a certain minimum detection limit (MDL) are censored. The proportions of such censored observations across datasets are not negligible. Ignoring the censoring by implementing some ad hoc methods such as replacing the censored values by MDL or MDL/2 leads to biased estimates of the overall spatial variability \citep{fridley2007data}. However, the studies of \cite{fridley2007data} were limited to a univariate spatial setting and to our knowledge, this important aspect of censoring has been completely ignored while modeling arsenic concentration in Bangladesh groundwater. 

Statistical inference for spatially distributed censored data has been studied quite extensively in the literature. Estimation and prediction methods have been developed based on the Expectation-Maximization (EM) algorithm \citep{militino1999analyzing, ordonez2018geostatistical}. To avoid computational challenges arising from censored likelihoods for correlated data, Monte Carlo approximations have been implemented under the classical \citep{stein1992prediction, rathbun2006spatial} and Bayesian paradigms \citep{kitanidis1986parameter, de2002bayesian, de2005bayesian, tadayon2017bayesian}. Finally, several data augmentation techniques have also been put forward to conveniently analyze spatially correlated censored data \citep{abrahamsen2001kriging, hopke2001multiple, fridley2007data, sedda2012imputing}. 

For many real datasets, it is often important to model multiple spatial processes jointly compared to modeling them independently or in a regression approach, where several variables are considered to be explanatory variables. {\color{black} Multivariate spatial models have been studied in a vast literature. \cite{mardia1993spatial} introduced separable cross-covariance functions in the context of spatiotemporal modeling and discussed a frequentist estimation procedure based on maximizing the underlying likelihood function. \cite{banerjee2002prediction} discussed a fully Bayesian implementation, and further, \cite{gelfand2003proper} proposed a separable model in the context of areal data. Apart from separable models, a popular multivariate modeling framework is the linear model of coregionalization \citep{wackernagel2003multivariate}. A nonstationary multivariate spatial model with spatially-varying coefficients has been introduced by \cite{gelfand2004nonstationary} and covariance convolution in this context has been proposed by \cite{majumdar2010generalized}. A multivariate non-Gaussian spatial model for skewed data has been proposed by \cite{hazra2019multivariate}. A detailed description of multivariate spatial models is in Chapter 7 of \cite{banerjee2015hierarchical}. Recently, \cite{kleiber2019model} proposed a model for large multivariate spatial datasets using a scalable multiresolution approach and \cite{guhaniyogi2019multivariate} proposed a metakriging approach in the same context. \cite{hazra2021large} discussed a spatial return level estimation approach for high-dimensional extremes based on a multivariate sparse Gaussian Markov random field.} 

When it comes to arsenic contamination analysis, \cite{lockwood2004analysis} suggested a Bayesian model for the joint distribution of seven groundwater elements, including arsenic in community water systems in the United States. \cite{guinness2014multivariate} and \cite{terres2018bayesian} studied the dependency between arsenic and other elements in soil samples from Clayton, North Carolina, USA under frequentist and Bayesian setups, respectively. However, as suggested by \cite{islam2000arsenic}, the concentrations of arsenic in Bangladesh groundwater are much higher compared to that in surface water or surface soil. Also, according to \cite{ohno2005arsenic}, there is some evidence of possible correlations between concentrations of arsenic and other elements in Bangladesh groundwater. As a result, a multivariate spatial model is required to analyze the joint spatial dependency among the elements in Bangladesh groundwater.


In this paper, we study the concentration of As, Ba, and Ca, in groundwater collected by the Bangladesh Water Development Board (BWDB) Water-Quality Monitoring network from 113 boreholes located throughout Bangladesh. Exploratory data analysis reveals that the concentrations of these elements are indeed correlated. For a significant proportion (18 out of 113) of the boreholes, arsenic concentration levels are below the MDL (0.5 $\mu$g) and they are left-censored. Therefore, we propose a joint multivariate hierarchical Bayesian spatial model with a separable covariance structure to capture the spatial distribution of arsenic concentration in Bangladesh groundwater, taking into account its dependency on other groundwater elements. Inference about model parameters is drawn based on an adaptive Markov Chain Monte Carlo (MCMC) sampling scheme, which is a combination of Gibbs sampling and random walk Metropolis-Hastings (M-H) steps. The proposed model easily accounts for the censoring in arsenic contamination, thereby avoiding any computational burden associated with multivariate likelihoods for censored observations. Based on the spatial maps obtained by fitting the proposed model, we also study the spatial variability of arsenic contamination across different divisions of Bangladesh.



The rest of the paper is organized as follows. In Section \ref{data}, the Bangladesh groundwater data are described in more details. Section \ref{methodology} presents the proposed multivariate spatial Bayesian model. Section \ref{computation} outlines the computational details for Bayesian inference. We perform some simulation studies in Section \ref{simulation} to assess model performance under different settings. In Section \ref{application}, the model results including maps of predicted arsenic concentrations over Bangladesh and the associated uncertainties are presented. Finally, Section \ref{discussion} concludes with a brief discussion of the presented methodology and potential future work.

\section{Data description and exploratory analysis}
\label{data}

The data used in this paper results from a national-scale survey of groundwater quality carried out at 113 boreholes from the Water-Quality Monitoring Network maintained by the Bangladesh Water Development Board (BWDB). The sites are located in all districts except three districts of the Chittagong Hill Tracts and Sunamganj in the northeast. One of the main aims of the investigation was to assess the scale of the groundwater arsenic problem to rapidly develop mitigation programs. A second aim was to increase the understanding of the origins and behavior of arsenic in Bangladesh groundwater. The data contains measurements of concentration (in $\mu$g/L) of arsenic in the groundwater, along with multiple other elements. Out of the As concentrations at the 113 boreholes, 95 are observed and 18 (15.9\%) are left-censored (that is, falling below an MDL which is fixed at 0.5 $\mu$g/L). The data is available at \url{https://www2.bgs.ac.uk/groundwater/health/arsenic/Bangladesh/data.html}. 

\begin{figure}[h]
\centering
\includegraphics[height = 0.4\linewidth]{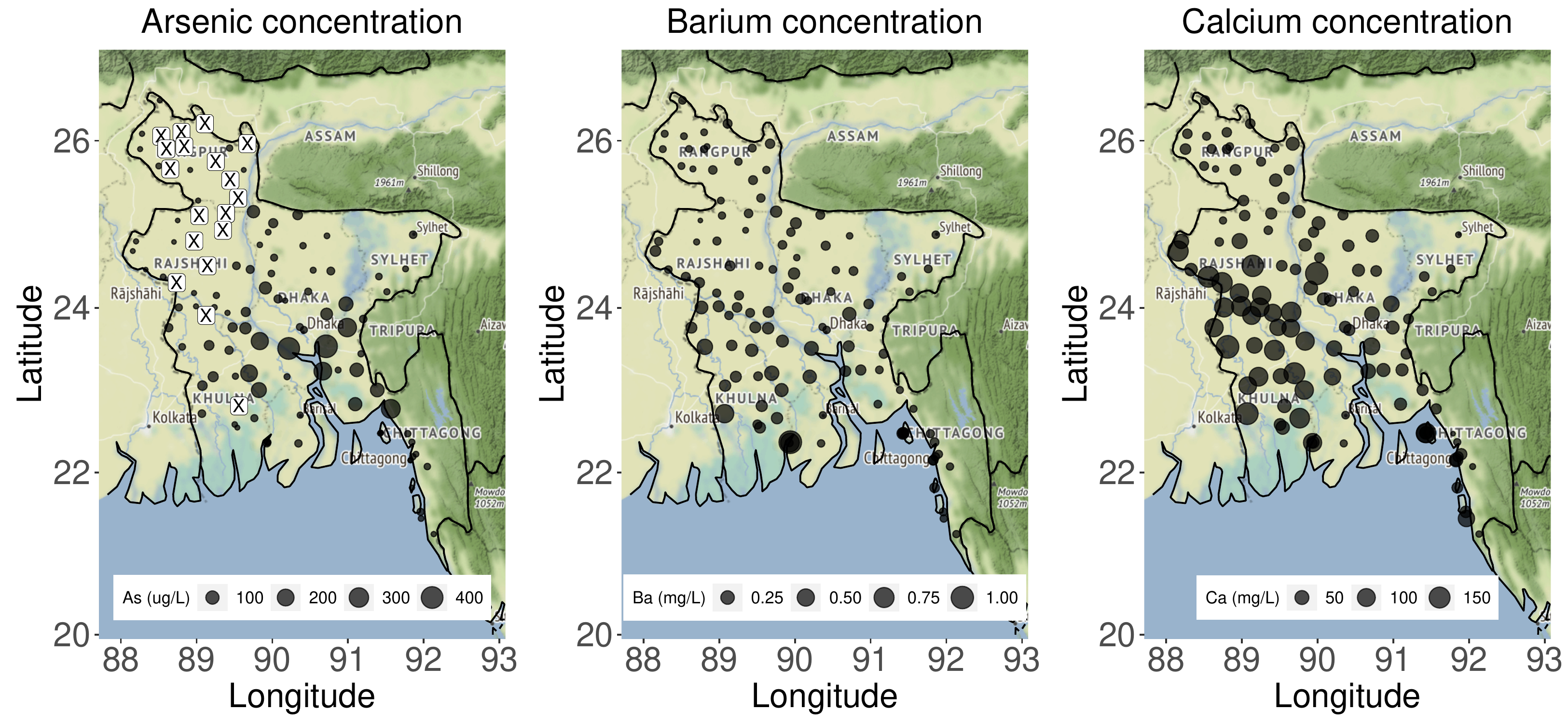}
\caption{Spatial maps showing the locations of the boreholes, and the corresponding concentrations of arsenic, barium, and calcium, in the groundwater. The `$\times$' symbols in the left panel show the locations where the arsenic concentrations are below the minimum detection limit (0.5 $\mu$g/L).}
\label{spatial_maps}
\end{figure}

We focus our analysis on groundwater concentrations of arsenic, barium and calcium. Figure \ref{spatial_maps} shows the locations of the boreholes (including the censored locations) across Bangladesh along with the concentrations of arsenic, barium and calcium.

\begin{figure}[h]
\centering
\includegraphics[width = \linewidth]{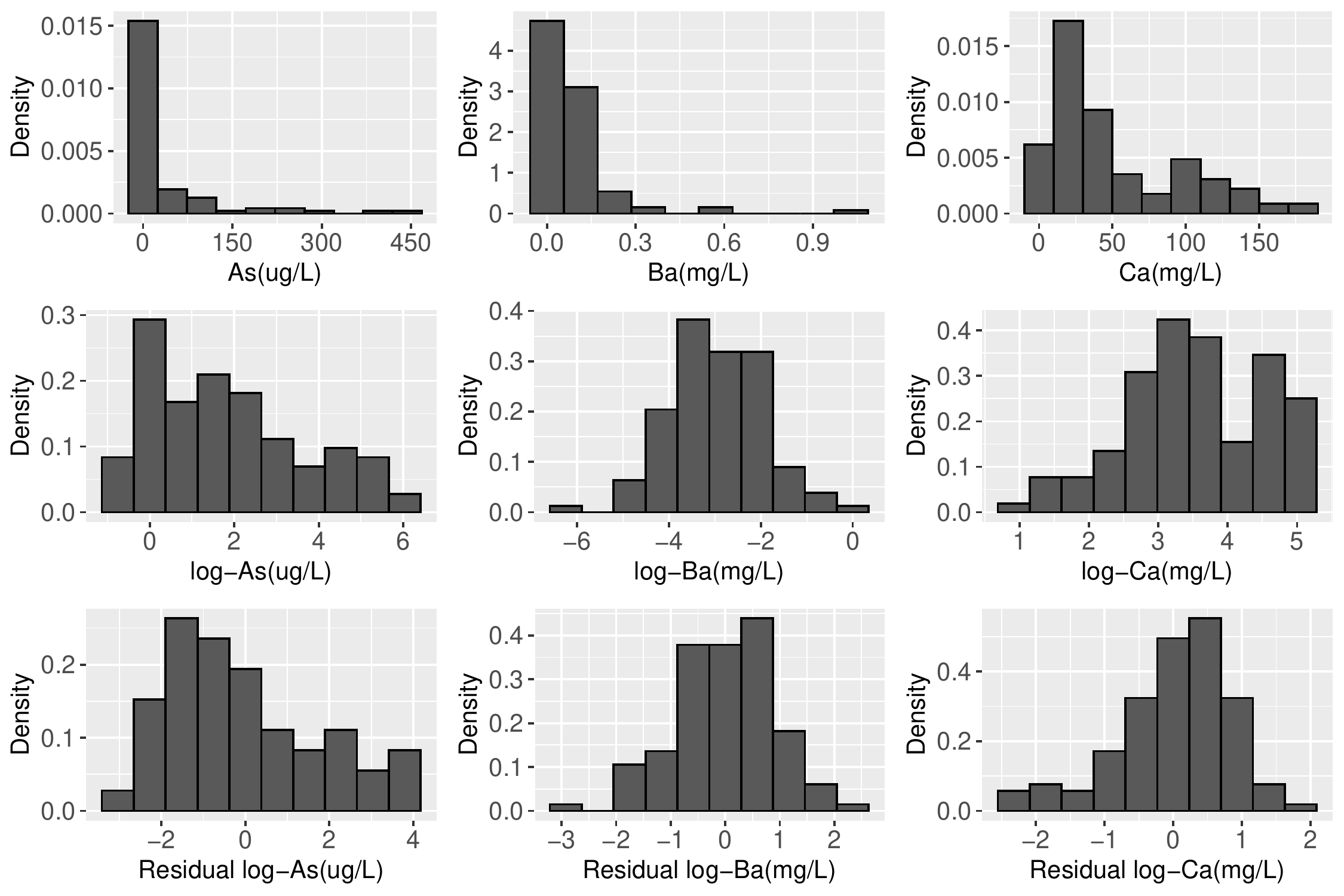}
\caption{First row: Histograms of arsenic, barium, and calcium concentrations. Second row: Histograms of the corresponding log-transformed concentrations. Third row: Histograms of the log-concentration residuals, after regressing on longitude and latitude.}
\label{histograms}
\end{figure}

\begin{figure}{}
\centering
\includegraphics[height = 0.4\linewidth]{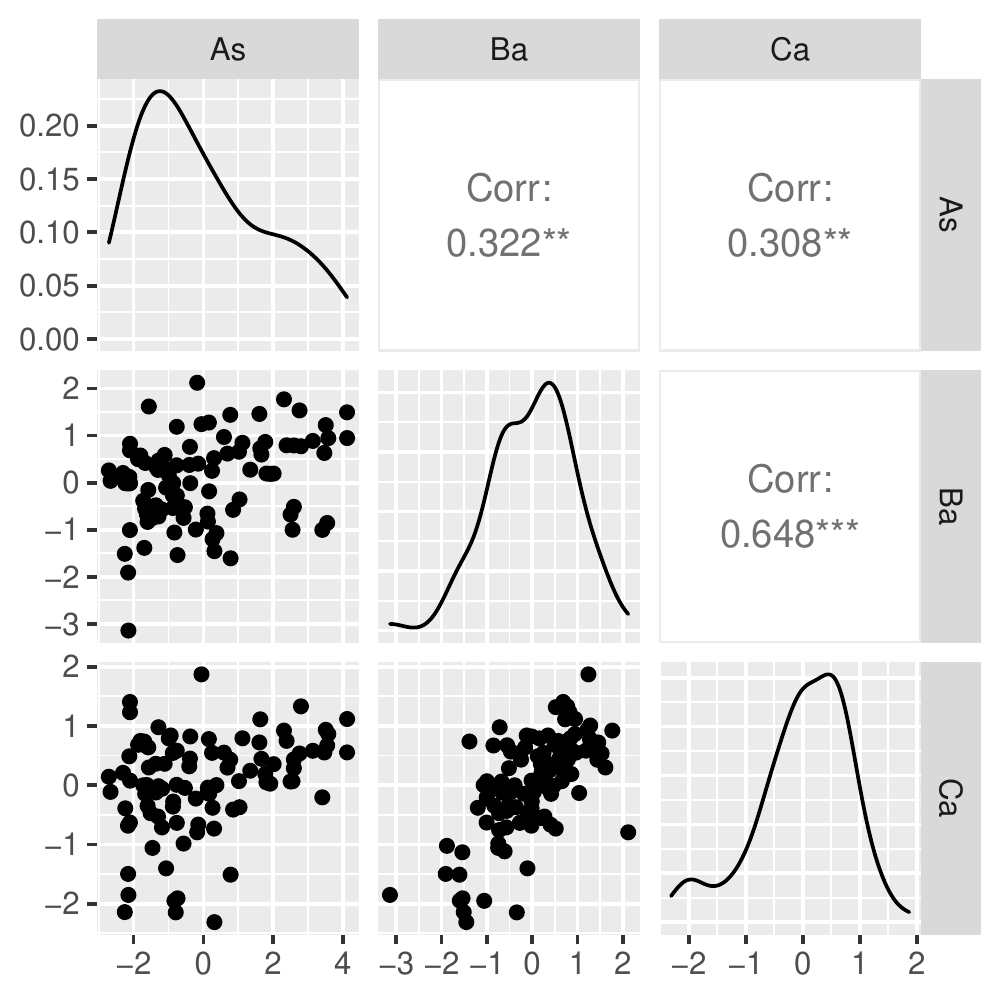}
\includegraphics[height = 0.4\linewidth]{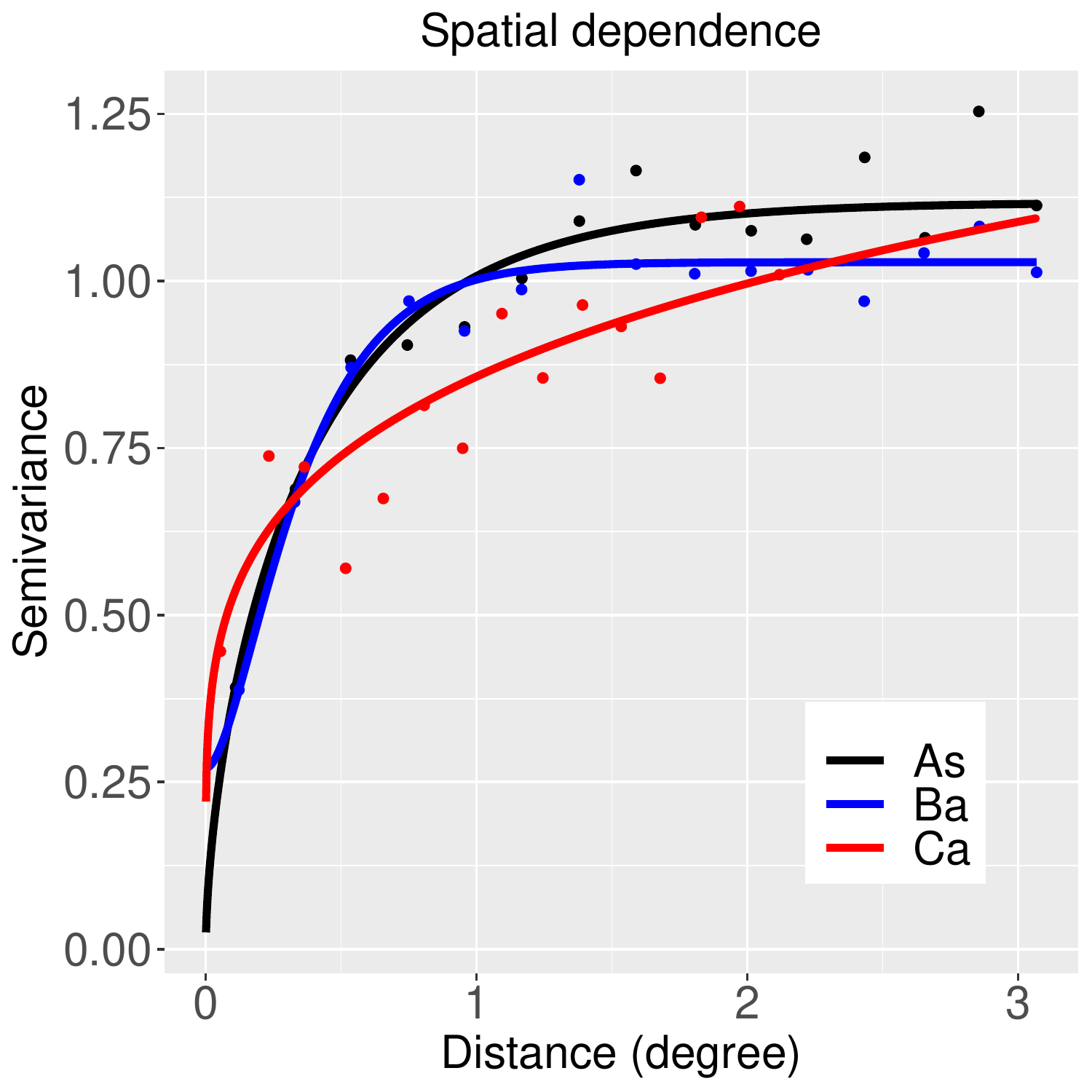}
\caption{Left: The scatter plots between pairs of As, Ba, and Ca log-concentration residuals (below the diagonal), the diagonals represent the kernel density estimates, and the upper diagonal elements denote the pairwise correlations. Right: Semivariograms of As, Ba, and Ca log-concentration residuals as functions of distance.}
\label{corplots}
\end{figure}

An initial exploratory analysis of the data shows that the distributions of arsenic, barium, and calcium concentrations are all right-skewed (see Figure \ref{histograms}, first row). Hence, the concentration measurements have been log-transformed to normalize the skewed distributions (see Figure \ref{histograms}, second row). We use longitude and latitude as covariates, and the third row of Figure \ref{histograms} displays the histograms of residuals obtained after fitting a simple linear regression to the log concentrations. Since our goal is to jointly model the log concentrations in the spatial domain, the dependencies among arsenic, barium, and calcium are displayed in the left panel of Figure \ref{corplots}. The left panel of Figure \ref{corplots} shows scatterplots between pairs of arsenic, barium, and calcium log-concentration residuals (after regressing on longitude and latitude), along with the pairwise correlation values. The diagonal elements show kernel density estimates of the distributions of the log-concentration residuals.

To visualize spatial correlations in the log-transformed concentration variables, we look at sample semivariograms for each variable. The sample semivariogram at distance $d$ is defined as 
$$
\widehat{\gamma}(d) = \frac{1}{2N(d)}\sum_{i = 1}^n \sum_{j = 1}^{i} w_{ij}(d)(Y(\bm{s}_i) - Y(\bm{s}_j))^2
$$
where $w_{ij}(d) = 1$ if $d_{ij} \in (d - h, d + h)$ and $w_{ij} = 0$ otherwise, $d_{ij}$ being the distance between $\bm{s}_i$ and $\bm{s}_j$. Also, $N(d)$ is the number of pairs with $w_{ij}(d) = 1$. As seen in the right panel of Figure \ref{corplots}, the sample variograms justify an exponential covariance structure for the stochastic component of our model and also the spatial ranges are reasonably similar, at least for arsenic and barium.

\section{Methodology}
\label{methodology}

In this paper, we present a joint multivariate spatial model using a hierarchical Bayesian framework to explain dependencies among concentrations of arsenic, barium, and calcium in Bangladesh groundwater. Our goal is to create spatial maps of arsenic, barium, and calcium concentrations and hence, our focus is on spatial prediction.


We denote the observation from the $p$-th variable at location $\bm{s}$ within the spatial domain of interest $\mathcal{D} \subset \mathbb{R}^2$ by $Y_p(\bm{s})$. For $p = 1, 2, \ldots, P$, we model $Y_p(\bm{s})$ as 
\begin{eqnarray} \label{main_model}
Y_p(\bm{s}) = \bm{X}(\bm{s})' \bm{\beta}^*_p + \varepsilon_p(\bm{s}) + \eta_p(\bm{s}),
\end{eqnarray}
where $\bm{X}(\bm{s}) = [X_1(\bm{s}), \ldots, X_Q(\bm{s})]'$ denotes the matrix of $Q$ covariates observed at location $\bm{s}$ and $\bm{\beta}^*_p = [\beta_{p,1}, \ldots, \beta_{p,Q}]'$. For our analysis we choose $\bm{X}(\bm{s}) = [1, \textrm{longitude}(\bm{s}), \textrm{latitude}(\bm{s})]'$. Also,  $\bm{\varepsilon}(\bm{s}) = [\varepsilon_1(\bm{s}), \ldots, \varepsilon_P(\bm{s})]'$ is assumed to be a multivariate spatial Gaussian process with separable correlation structure. In particular, $\bm{\varepsilon}(\bm{s}) \sim \textrm{Normal}_P(\bm{0}, r\bm{\Sigma})$ at every location $\bm{s}$, and for each $p$, $\varepsilon_p(\cdot)$ exhibits exponential spatial correlation, that is, 
\begin{eqnarray}\label{cor_structure}
 \textrm{Cor}[\varepsilon_{p_1}(\bm{s}_i), \varepsilon_{p_2}(\bm{s}_j)] = r \Sigma_{p_1, p_2} \exp[-\lVert \bm{s}_i - \bm{s}_j \rVert / \phi].
\end{eqnarray} 
Here $\lVert \bm{s}_i - \bm{s}_j \rVert$ denotes the geodesic distance (in kilometers) implemented in \texttt{rdist.earth} function in the \texttt{R} package \texttt{fields}, and $\Sigma_{p_1, p_2}$ denotes the $(p_1, p_2)$-th element of $\bm{\Sigma}$. In addition, $\bm{\eta}(\bm{s}) = [\eta_1(\bm{s}), \ldots, \eta_P(\bm{s})]'$ denotes the multivariate nugget effect with $\bm{\eta}(\bm{s}) \sim \textrm{Normal}_P(\bm{0}, (1-r)\bm{\Sigma})$. Here $r \in [0, 1]$ is the ratio of spatial to total variation. The multivariate nugget term tackles the censoring in arsenic log-concentration, thereby circumnavigating computational burden occurring due to censored likelihoods \citep{hazra2018semiparametric, yadav2019spatial, zhang2021hierarchical}. 



We denote the observation vector at location $\bm{s}$ by $\bm{Y}(\bm{s}) = [Y_1(\bm{s}), \ldots, Y_P(\bm{s})]'$. For observation locations $\mathcal{S} = \lbrace \bm{s}_1, \ldots, \bm{s}_N \rbrace \subset \mathcal{D}$, define $\bm{Y} = [\bm{Y}(\bm{s}_1)', \ldots, \bm{Y}(\bm{s}_N)']'$,  $\bm{\varepsilon} = [\bm{\varepsilon}(\bm{s}_1)', \ldots, \bm{\varepsilon}(\bm{s}_N)']'$ and $\bm{\eta} = [\bm{\eta}(\bm{s}_1)', \ldots, \bm{\eta}(\bm{s}_N)']'$. Also, let $\bm{X}$ denote the $N\times Q$-dimensional design matrix with its $i$-th row being $\bm{X}(\bm{s}_i), i=1, \ldots, N$ and $\bm{\beta}$ denote the full vector of regression coefficients, $\bm{\beta} = (\beta_{1,1}, \ldots, \beta_{P,1}, \beta_{1,2}, \ldots, \beta_{P,2}, \ldots, \beta_{1,Q}, \ldots, \beta_{P,Q})'$. Using the vector-matrix notations, the full model can be written as
$$\bm{Y} = [\bm{X} \otimes \bm{I}_P] \bm{\beta} + \bm{\varepsilon} + \bm{\eta},$$
where
$\bm{\varepsilon} \sim \textrm{Normal}_{NP}(\bm{0}, r \bm{\Sigma}_{\mathcal{S}} \otimes \bm{\Sigma})$ and $\bm{\eta} \sim \textrm{Normal}_{NP}(\bm{0}, (1 - r) \bm{I}_{N} \otimes \bm{\Sigma})$. Here $\bm{\Sigma}_{\mathcal{S}}$ denotes the $N\times N$-dimensional correlation matrix between the spatial locations $\{ \bm{s}_1, \ldots, \bm{s}_N\}$ induced by the correlation structure (\ref{cor_structure}).


The joint distribution of $\bm{Y}$ after marginalizing over $\bm{\varepsilon}$ is
\begin{eqnarray}\label{model}
\bm{Y} \sim \textrm{Normal}_{NP}([\bm{X} \otimes \bm{I}_P] \bm{\beta}, [r \bm{\Sigma}_{\mathcal{S}} + (1 - r) \bm{I}_{N}] \otimes \bm{\Sigma}).
\end{eqnarray}
Thus, the final process after marginalization indeed has a separable covariance structure \citep{gelfand2003proper}.

Motivated by the dataset considered, we assume $Y_1(\cdot)$ is left-censored at the spatial locations $\mathcal{S}^{(c)} = \{ \bm{s}^{(c)}_1, \ldots, \bm{s}^{(c)}_{N_c} \} \subset \mathcal{S}$ and the censoring level is $u$. For the sake of simplicity, we consider the same type of censoring. However, a similar approach can be applied if multiple variables have censoring, possibly at different spatial locations. Define the censoring indicator $\delta(\bm{s})$ as 
$$
\delta(\bm{s}) = \twopartdef{1, }{\text{if } Y_1(\bm{s}) \text{ is censored at location } \bm{s}}{0, }{\text{otherwise}.}
$$
and the vector of censored observations as $$\bm{v} = [ Y_1(\bm{s}_i): \delta(\bm{s}_i) = 1]' \equiv [Y_1(\bm{s}^{(c)}_1), \ldots, Y_1(\bm{s}^{(c)}_{N_c})]'.$$
Then, for censored spatial data, the likelihood is given by
$$
\mathcal{L}(\bm{\theta}) = \int_{\bm{v} \leq u} f_{\textrm{Normal}_{NP}}(\bm{y}; [\bm{X} \otimes \bm{I}_P] \bm{\beta}, [r \bm{\Sigma}_{\mathcal{S}} + (1 - r) \bm{I}_{N}] \otimes \bm{\Sigma}) \hspace{0.05cm} d\bm{v},
$$
where the intergral is over the censored region $\lbrace \bm{y}: y_1(\bm{s}_i) \leq u ~\textrm{if}~ \bm{s}_i \in  \mathcal{S}^{(c)} \rbrace$ and $f_{\textrm{Normal}_n}(\cdot; \bm{\mu}, \bm{\Sigma})$ denotes the $n$-variate normal density with mean $\bm{\mu}$ and covariance matrix $\bm{\Sigma}$.


\subsection{Prediction}
\label{prediction}

We denote the prediction locations by $\mathcal{S}^{(0)} = \lbrace \bm{s}^{(0)}_1, \ldots, \bm{s}^{(0)}_M \rbrace \subset \mathcal{D}$, and define $\bm{Y}^{(0)} = [\bm{Y}(\bm{s}^{(0)}_1)', \ldots, \bm{Y}(\bm{s}^{(0)}_M)' ]'$, $\bm{\varepsilon}^{(0)} = [\bm{\varepsilon}(\bm{s}_1^{(0)})', \ldots, \bm{\varepsilon}(\bm{s}_M^{(0)})']'$ and $\bm{\eta}^{(0)} = [\bm{\eta}(\bm{s}_1^{(0)})', \ldots, \bm{\eta}(\bm{s}_N^{(0)})']'$. Also, $\bm{X}^{(0)}$ denotes the $M\times Q$-dimensional design matrix with its $i_0$-th row being $\bm{X}(\bm{s}^{(0)}_i), i_0=1, \ldots, M$. Denoting the exponential correlation matrix between the prediction locations $\mathcal{S}^{(0)}$ by $\bm{\Sigma}_{\mathcal{S}}^{(0, 0)}$, the correlation matrix between the locations $\mathcal{S}^{(0)}$ and $\mathcal{S}$ by $\bm{\Sigma}_{\mathcal{S}}^{(0, \cdot)}$ and its transpose by $\bm{\Sigma}_{\mathcal{S}}^{(\cdot, 0)}$, the conditional distribution of $\bm{\varepsilon}^{(0)}$ given $\bm{\varepsilon}$ is
\begin{eqnarray}
\nonumber \bm{\varepsilon}^{(0)} \vert \bm{\varepsilon} \sim \textrm{Normal}_{MP} \left( \left[ \bm{\Sigma}_{\mathcal{S}}^{(0, \cdot)} \bm{\Sigma}_{\mathcal{S}}^{-1} \otimes \bm{I}_P \right] \bm{\varepsilon}, r \left[\bm{\Sigma}_{\mathcal{S}}^{(0, 0)} - \bm{\Sigma}_{\mathcal{S}}^{(0, \cdot)} \bm{\Sigma}_{\mathcal{S}}^{-1} \bm{\Sigma}_{\mathcal{S}}^{(\cdot, 0)}\right] \otimes \bm{\Sigma} \right)
\end{eqnarray}
and the conditional distribution of $\bm{Y}^{(0)}$ given $\bm{Y}$ and $\bm{\varepsilon}^{(0)}$ is
\begin{eqnarray}
\nonumber \bm{Y}^{(0)} \vert \bm{Y}, \bm{\varepsilon}^{(0)} \sim \textrm{Normal}_{MP} \left( [\bm{X}^{(0)} \otimes \bm{I}_P] \bm{\beta} + \bm{\varepsilon}^{(0)} , (1-r) \bm{I}_M \otimes \bm{\Sigma} \right).
\end{eqnarray}

The conditional distribution of $\bm{Y}^{(0)}$ given only $\bm{Y}$ is obtained by marginalizing with respect to the latent Gaussian process $\bm{\varepsilon}(\cdot)$. For the real data application, we choose prediction locations at a resolution of $0.15^\circ \times 0.15^\circ$ across Bangladesh which leads to $M=526$ grid cells.

\section{{\color{black} Posterior inference and computational details}}
\label{computation}

We draw inference about the model parameters based on Markov chain Monte Carlo (MCMC) sampling, implemented in \texttt{R}. As the computation is dependent on the choice of priors for the model parameters, we specify the priors first. We select conjugate priors when possible and update them using Gibbs sampling. For some parameters, conjugate prior distributions do not exist. In such situations, we use random walk Metropolis-Hastings steps to update the parameters. We tune the candidate distributions in Metropolis-Hastings steps during the burn-in period so that the acceptance rate during the post-burn-in period remains between 0.3 and 0.5. 

In our fully Bayesian analysis, the latent multivariate process $\bm{\varepsilon}(\cdot)$, the censored observations and the observations at the prediction locations $\bm{Y}^{(0)}$ are also treated as parameters. The set of parameters and hyper-parameters in the model are $$\Theta = \left\lbrace \bm{\beta}, \bm{\Sigma}, \phi, r, {\bm{\varepsilon}}, Y_1\left(\bm{s}^{(c)}_1\right), \ldots, Y_1\left(\bm{s}^{(c)}_{N_c}\right), \bm{Y}^{(0)} \right \rbrace.$$ The MCMC steps for updating the parameters in $\Theta$ are as follows. Corresponding to a parameter (or a set of parameters), by $rest$, we mean the data, all the parameters and hyperparameters in $\Theta$ except that parameter (or that set of parameters).

For the vector of regression coefficients $\bm{\beta}$, we consider less-informative conjugate prior $\bm{\beta} \sim \textrm{Normal}_{PQ}(\bm{0}, 100^2 \bm{I}_Q \otimes \bm{\Sigma})$. The full posterior distribution of $\bm{\beta}$ is multivariate normal and is given by $\bm\beta | rest \sim \textrm{Normal}_{PQ}(\bm{\mu}_{\bm{\beta}}^\ast, \bm{\Sigma}_{\bm{\beta}}^\ast)$, where
\begin{eqnarray}
\nonumber && \bm{\Sigma}_{\bm{\beta}}^\ast = \left[\frac{1}{1-r} {\bm{X}}'{\bm{X}} + 100^{-2} \bm{I}_Q \right]^{-1} \otimes \bm{\Sigma},\\
\nonumber && \bm{\mu}_{\bm{\beta}}^\ast = \left[\left( \left[\frac{1}{1-r} {\bm{X}}'{\bm{X}} + 100^{-2} \bm{I}_Q \right]^{-1} \frac{1}{1-r} {\bm{X}}' \right) \otimes \bm{I}_P  \right] \left({\bm{Y}} - {\bm{\varepsilon}} \right),
\end{eqnarray}
and hence, $\bm{\beta}$ is updated using Gibbs sampling. Due to the choice of the separable covariance structure of the prior for $\bm{\beta}$, the full conditional posterior covariance matrix is also separable. 



Now, let $\bm{B}$ denote the $(Q\times P)$-dimensional matrix obtained by stacking $\bm{\beta}_q, q=1, \ldots, Q$ across the rows and $\bm{Y}^*$ and $\bm{E}$ denote the $(N \times P)$-dimensional matrices obtained by stacking $\bm{Y}(\bm{s}_1), \ldots, \bm{Y}(\bm{s}_N)$, and $\bm{\epsilon}(\bm{s}_1), \ldots, \bm{\epsilon}(\bm{s}_N)$ across the rows, respectively. For $\bm{\Sigma}$, we assume the non-informative conjugate prior $\bm{\Sigma} \sim \textrm{Inverse-Wishart}(0.01, 0.01 \bm{I}_P)$. The full conditional posterior density is $\bm{\Sigma}| rest \sim \textrm{Inverse-Wishart}(\nu, \bm{\Psi})$, where
\begin{eqnarray}
\nonumber \nu &=& 0.01 + 2N + 2M + Q, \\
\nonumber \bm{\Psi} &=& 0.01 \bm{I}_P + (\bm{Y}^* - {\bm{X}}\bm{B} - \bm{E})'(\bm{Y}^* - {\bm{X}}\bm{B} - \bm{E}) / (1 - r) + \bm{E}'\bm{\Sigma}_{\mathcal{S}}^{-1}\bm{E}/r + 100^{-2} \bm{B}'\bm{B},
	\end{eqnarray}
and hence, $\bm{\Sigma}$ is also updated using Gibbs sampling.

For the range parameter $\phi$ in (\ref{cor_structure}), we consider the prior $\phi \sim \textrm{Uniform}(0, 0.5\Delta)$, where $\Delta$ is the largest geodesic distance between two data locations. Suppose $\phi^{(m)}$ denotes the $m$-th MCMC sample corresponding to $\phi$. Considering a logit transformation, we obtain $\phi^{*(m)} \in \mathbb{R}$ from $\phi^{(m)}$ and simulate $\phi^{*(c)} \sim \textrm{Normal}( \phi^{*(m)}, s_{\phi}^2)$, where $s_{\phi}$ is the standard deviation of the candidate normal distribution. Subsequently, using an inverse-logit transformation, we obtain $\phi^{(c)}$ from $\phi^{*(c)}$ and consider $\phi^{(c)}$ to be a candidate from the posterior distribution of $\phi$. Let $\bm{\Sigma}_{\mathcal{S}}^{(m)}$ and $\bm{\Sigma}_{\mathcal{S}}^{(c)}$ denote the spatial correlation matrices corresponding to $\mathcal{S}$, with $\phi=\phi^{(m)}$ and $\phi = \phi^{(c)}$, respectively. The acceptance ratio is 
\begin{eqnarray}
\nonumber \mathcal{R} &=& \frac{ f_{\textrm{Normal}_{NP}}\left({\bm{\varepsilon}}; \bm{0}, r \bm{\Sigma}_{\mathcal{S}}^{(c)} \otimes \bm{\Sigma} \right) }{ f_{\textrm{Normal}_{NP}}\left({\bm{\varepsilon}}; \bm{0}, r \bm{\Sigma}_{\mathcal{S}}^{(m)} \otimes \bm{\Sigma} \right)} \times \frac{\phi^{(c)} \left(0.5\Delta - \phi^{(c)}\right)}{\phi^{(m)} \left(0.5\Delta - \phi^{(m)}\right)}.
\end{eqnarray}
The candidate is accepted with probability $min \lbrace \mathcal{R},1 \rbrace$.


For $r$, the ratio of spatial to total variation, we consider the prior $r \sim \textrm{Uniform}(0, 1)$. Suppose $r^{(m)}$ denotes the $m$-th MCMC sample from $r$. We simulate a candidate sample $r^{(c)}$ from $r^{(m)}$ following a procedure similar to simulating $\phi^{(c)}$ from $\phi^{(m)}$. The Metropolis-Hastings acceptance ratio is 
\begin{eqnarray}
\nonumber \mathcal{R} &=& \frac{f_{\textrm{Normal}_{NP}}\left({\bm{Y}}; [{\bm{X}} \otimes \bm{I}_P] \bm{\beta} + {\bm{\varepsilon}}, (1 - r^{(c)}) \bm{I}_{N} \otimes \bm{\Sigma} \right)}{f_{\textrm{Normal}_{NP}}\left({\bm{Y}}; [{\bm{X}} \otimes \bm{I}_P] \bm{\beta} + {\bm{\varepsilon}}, (1 - r^{(m)}) \bm{I}_{N} \otimes \bm{\Sigma} \right)} \\
\nonumber && \times \frac{ f_{\textrm{Normal}_{NP}}\left({\bm{\varepsilon}}; \bm{0}, r^{(c)} \bm{\Sigma}_{\mathcal{S}} \otimes \bm{\Sigma} \right) }{ f_{\textrm{Normal}_{NP}}\left({\bm{\varepsilon}}; \bm{0}, r^{(m)} \bm{\Sigma}_{\mathcal{S}} \otimes \bm{\Sigma} \right)} \times \frac{r^{(c)}  \left(1 - r^{(c)}\right)}{r^{(m)} \left(1 - r^{(m)}\right)}
\end{eqnarray}
and the candidate $r^{(c)}$ is accepted with probability $min \lbrace \mathcal{R},1 \rbrace$.


The unconditional distribution of ${\bm{\varepsilon}}$ is ${\bm{\varepsilon}} \sim \textrm{Normal}_{NP}(\bm{0}, r \bm{\Sigma}_{\mathcal{S}} \otimes \bm{\Sigma})$. The full conditional posterior distribution of ${\bm{\varepsilon}}$ is ${\bm{\varepsilon}} | rest \sim \textrm{Normal}_{NP}(\bm{\mu}^\ast_{{\bm{\varepsilon}}}, \bm{\Sigma}^\ast_{{\bm{\varepsilon}}})$, where
\begin{eqnarray}
\nonumber && \bm{\Sigma}^\ast_{{\bm{\varepsilon}}} = \left[(1 - r)^{-1} \bm{I}_{N} + r^{-1} \bm{\Sigma}_{\mathcal{S}}^{-1} \right]^{-1} \otimes \bm{\Sigma},\\
\nonumber && \bm{\mu}^\ast_{{\bm{\varepsilon}}} = \left[ (1 - r)^{-1}\left[(1 - r)^{-1} \bm{I}_{N} + r^{-1} \bm{\Sigma}_{\mathcal{S}}^{-1} \right]^{-1} \otimes \bm{I}_P \right] \left({\bm{Y}} - [{\bm{X}} \otimes \bm{I}_P] \bm{\beta}\right).
\end{eqnarray}


Additional to the model parameters and the latent Gaussian process $\bm{\varepsilon}(\cdot)$, the observations $Y_1(\bm{s}^{(c)}_1), \ldots, Y_1(\bm{s}^{(c)}_{N_c})$ are left-censored at $u$. Within MCMC, we need to impute the censored values at every iteration. They are updated independently in a similar way and hence, without loss of generality we consider updating $Y_1(\bm{s}^{(c)}_1)$. Define $\bm{Y}^{(-1)}(\bm{s}^{(c)}_1) = [Y_2(\bm{s}^{(c)}_1), \ldots, Y_P(\bm{s}^{(c)}_1)]'$ and hence, $\bm{Y}(\bm{s}^{(c)}_1) = [Y_1(\bm{s}^{(c)}_1), \bm{Y}^{(-1)}(\bm{s}^{(c)}_1)']'$. {\color{black} Let the unconditional mean of $\bm{Y}(\bm{s}^{(c)}_1)$ be denoted by $\bm{\mu}(\bm{s}^{(c)}_1) = \bm{B}'\bm{X}(\bm{s}^{(c)}_1)$ and $\bm{\mu}(\bm{s}^{(c)}_1) = [\mu_1(\bm{s}^{(c)}_1), \bm{\mu}^{(-1)}(\bm{s}^{(c)}_1)']'$, where $\bm{\mu}^{(-1)}(\bm{s}^{(c)}_1) = [\mu_2(\bm{s}^{(c)}_1), \ldots, \mu_P(\bm{s}^{(c)}_1)]'$.} Similarly, $\bm{\varepsilon}^{(-1)}(\bm{s}^{(c)}_1) = [\varepsilon_2(\bm{s}^{(c)}_1), \ldots, \varepsilon_P(\bm{s}^{(c)}_1)]'$ and hence, $\bm{\varepsilon}(\bm{s}^{(c)}_1) = [\varepsilon_1(\bm{s}^{(c)}_1), \bm{\varepsilon}^{(-1)}(\bm{s}^{(c)}_1)']'$. Further, denoting the $(1,1)$-th element of $\bm{\Sigma}$ by $\Sigma_{1,1}$, the rest of the first column by $\bm{\Sigma}_{-1, 1}$, the rest of the first row by $\bm{\Sigma}_{1, -1}$ and the matrix without the first row and first column by $\bm{\Sigma}_{-1,-1}$, the full conditional distribution of $Y_1(\bm{s}^{(c)}_1)$ is 
\begin{eqnarray}
\nonumber && Y_1(\bm{s}^{(c)}_1) | rest \sim \textrm{Truncated-Normal}_{(-\infty, u)}\left( \mu^*_{Y_1(\bm{s}^{(c)}_1)}, \sigma^{2*}_{Y_1(\bm{s}^{(c)}_1)} \right),~~~\textrm{where}\\
\nonumber && \mu^*_{Y_1(\bm{s}^{(c)}_1)} = \left[\mu_1(\bm{s}^{(c)}_1) + \varepsilon_1(\bm{s}^{(c)}_1) \right] + \bm{\Sigma}_{1, -1} \bm{\Sigma}^{-1}_{-1,-1} \left[ \bm{Y}^{(-1)}(\bm{s}^{(c)}_1) - \bm{\mu}^{(-1)}(\bm{s}^{(c)}_1) -  \bm{\varepsilon}^{(-1)}(\bm{s}^{(c)}_1) \right], \\
\nonumber && \sigma^{2*}_{Y_1(\bm{s}^{(c)}_1)} = (1-r) \left[ \Sigma_{1,1} - \bm{\Sigma}_{1, -1} \bm{\Sigma}^{-1}_{-1,-1} \bm{\Sigma}_{-1, 1} \right].
\end{eqnarray}

Finally, we simulate $\bm{Y}^{(0)}$, the observed multivariate spatial field at the prediction locations $\mathcal{S}^{(0)}$ following Section \ref{prediction}.  


For our data application, we run the MCMC chain for 70,000 iterations and discard first 20,000 iterations as burn-in. The post-burn-in samples are then thinned by keeping one in each five samples. Thus, we draw inference based on 10,000 post-burn-in samples. Convergence of the chains is monitored by trace plots, as displayed in Figure \ref{trace_plots}. The computing time for the Bangladesh contamination dataset is 62 minutes on a single core of a desktop with Intel Xeon CPU E5-2680 2.40 GHz processor and 128 GB RAM.





\section{Simulation studies}
\label{simulation}

In this section, we perform some simulation studies to determine the performance of our model in terms of spatial prediction while imputing censored values in randomly generated datasets. For simplicity, we assume that the spatial process is bivariate, where the first variable is censored below a certain data percentile point and the second variable does not have any censoring. We simulate 100 datasets over 256 grid cells $\mathcal{S}^* = \{(i,j): i, j\in \{ 0, \ldots, 15 \}\}$ within a $[0, 15]^2$ spatial domain. We divide each dataset into training and test sets. We randomly choose 50 spatial locations for the test set. Within the training set, we consider two different levels of censoring (denoted by L1 and L2) for the first variable:
\begin{itemize}
    \item[L1] Low censoring: The MDL is at the $15^{\mbox{th}}$ percentile point of observations. 
    \item[L2] High censoring: The MDL is at the $45^{\mbox{th}}$ percentile point of observations. 
\end{itemize}

For each of these two levels of censoring, we implement our proposed model under three different settings (denoted by S1, S2, and S3):

\begin{itemize}
    \item[S1] We fix the censored observations at MDL and implement the multivariate spatial model as in (\ref{main_model}). This does not require any imputation of the censored observations.
    \item[S2] We ignore the spatial locations where the observations are censored and implement the multivariate spatial model as in (\ref{main_model}). Once again, this does not require any imputation of the censored observations.
    \item[S3] We fit the full proposed model, that is, we treat the observations below MDL as censored observations and implement the multivariate spatial model as in (\ref{main_model}) along with imputation of the censored observations.
\end{itemize}

We consider a similar design matrix as in (\ref{main_model}), in which the second and the third columns are centered and scaled to have mean zero and variance one.

For simulating the datasets, we assume the regression coefficients for the two variables to be $\bm{\beta}^*_{1} = [4, 0, 0]'$ and $\bm{\beta}^*_{2} = [6, 0, 0]'$ respectively. We also assume that the diagonal elements of $\bm{\Sigma}$ are 2 and the off-diagonals are 1, thereby setting the correlation between the two variables to be 0.5. While we choose geodesic distance for the data application as mentioned in Section \ref{methodology}, geodesic distance is not meaningful in this scenario and hence, we replace it with Euclidean distance in this section. The range parameter of the spatial exponential correlation is chosen to be $\phi = 2.5$ and the ratio of partial sill to total variation is chosen to be $r=0.8$. The prior distributions for $\bm{\beta}$, $\bm{\Sigma}$, and $r$ as described in Section \ref{computation} remain unchanged in the simulation study. However, for the range parameter we assume $\phi \sim \textrm{Uniform}(0, 0.25\Delta^*)$, where  $\Delta^*$ is the largest Euclidean distance between two data locations in $\mathcal{S}^*$. 

We compare the performances of the model under different combinations of L1 and L2 with S1, S2, and S3 in terms of root mean squared error (RMSE) while estimating model parameters and in terms of continuous rank probability score (CRPS) while predicting observations in the test set. Smaller values of both RMSE and CRPS are preferred. 

Table \ref{table1} displays the average RMSE while estimating the model parameters under different combinations of censoring levels and settings based on 100 simulated datasets. The corresponding standard errors are given in parentheses. When the level of censoring in the data is low, the parameters estimates obtained from models under S1 and S3 are comparable. However, the estimates, especially for the covariance parameters, are unreliable if the spatial locations with censored observations are ignored completely. On the other hand, when the level of censoring in the data is high, the final model along with imputation of the censored observations (S3) performs much better compared to models under S1 and S2, especially while estimating the covariance parameters.

\begin{table}[ht]
\caption{Average RMSE in estimation of model parameters under different censoring levels L1 (low-censoring) and L2 (high-censoring) and different settings S1, S2 and S3 based on 100 simulated datasets. The values within the parentheses are the corresponding standard errors. A smaller value of RMSE indicates better performance in parameter estimation.}
\centering
\begin{tabular}{cccc}
\hline
\multicolumn{4}{c}{L1: Low-censoring} \\
  \hline
Parameter & S1 & S2 & S3 \\ 
  \hline
$\beta_{1,1}$ & 0.503 (0.018) & 0.547 (0.022) & 0.543 (0.018) \\ 
$\beta_{2,1}$ & 0.526 (0.017) & 0.521 (0.018) & 0.532 (0.017) \\ 
$\beta_{1,2}$ & 0.345 (0.009) & 0.318 (0.008) & 0.400 (0.011) \\ 
$\beta_{2,2}$ & 0.399 (0.013) & 0.383 (0.013) & 0.404 (0.013) \\ 
$\beta_{1,3}$ & 0.352 (0.011) & 0.319 (0.010) & 0.399 (0.013) \\ 
$\beta_{2,3}$ & 0.366 (0.010) & 0.354 (0.009) & 0.373 (0.010) \\ 
$\Sigma_{1,1}$ & 0.554 (0.015) & 0.651 (0.018) & 0.521 (0.017) \\ 
$\Sigma_{2,2}$ & 0.491 (0.018) & 0.481 (0.013) & 0.511 (0.019) \\ 
$\Sigma_{1,2}$ & 0.295 (0.008) & 0.349 (0.010) & 0.320 (0.011) \\ 
$\phi$ & 1.081 (0.024) & 1.146 (0.022) & 1.090 (0.025) \\ 
$r$ & 0.092 (0.004) & 0.120 (0.006) & 0.091 (0.003) \\ 
   \hline
\multicolumn{4}{c}{L2: High-censoring} \\
  \hline
Parameter & S1 & S2 & S3 \\ 
\hline
$\beta_{1,1}$ & 0.575 (0.028) & 0.888 (0.038) & 0.559 (0.017) \\ 
$\beta_{2,1}$ & 0.513 (0.017) & 0.613 (0.028) & 0.542 (0.019) \\ 
$\beta_{1,2}$ & 0.241 (0.006) & 0.243 (0.006) & 0.411 (0.012) \\ 
$\beta_{2,2}$ & 0.391 (0.013) & 0.360 (0.012) & 0.404 (0.013) \\ 
$\beta_{1,3}$ & 0.244 (0.008) & 0.241 (0.007) & 0.415 (0.013) \\ 
$\beta_{2,3}$ & 0.356 (0.010) & 0.340 (0.010) & 0.374 (0.009) \\ 
$\Sigma_{1,1}$ & 1.163 (0.016) & 1.119 (0.018) & 0.596 (0.022) \\ 
$\Sigma_{2,2}$ & 0.463 (0.015) & 0.560 (0.016) & 0.521 (0.021) \\ 
$\Sigma_{1,2}$ & 0.477 (0.011) & 0.559 (0.013) & 0.351 (0.014) \\ 
$\phi$ & 1.096 (0.022) & 1.228 (0.020) & 1.087 (0.025) \\ 
$r$ & 0.105 (0.005) & 0.207 (0.010) & 0.097 (0.004) \\ 
   \hline
\end{tabular}
\label{table1}
\end{table}

Because our primary goal is predicting observations at new locations to create spatial maps, we use the continuous rank probability score \citep[CRPS;][]{matheson1976scoring, hersbach2000decomposition, gneiting2007strictly} to assess how well the model performs in terms of spatial prediction under the different scenarios. For a single test sample $y$, the CRPS is defined as $${\rm CRPS}(y,F)=\int_{-\infty}^{\infty} \left\{F(x) - \mathbb{I}_{\lbrace y \leq x \rbrace} \right\}^2 {\rm d}x,$$ where $F$ is the posterior predictive distribution function. We report the results by averaging values over the test set.

Table \ref{table2} displays the average CRPS while assessing spatial prediction under different combinations of censoring levels and settings based on 100 simulated datasets. The corresponding standard errors are mentioned in parentheses. Here, Variable (denoted by V) 1 includes censoring and we note that the final model along with imputation of the censored observations (S3) performs significantly better in spatial prediction for Variable 1 compared to models under settings S1 or S2. Also, the higher the level of censoring, the worse are the performance of models under S1 or S2. Thus, we can conclude that a full model with the imputation of censored data is preferred while modeling multivariate spatial censored data.

{\color{black} Table \ref{table3} reports the empirical coverage probabilities of the 90\% and 95\% prediction intervals (averaged across the prediction locations) under different combinations of censoring levels and settings based on the same 100 simulated datasets. The corresponding standard errors are mentioned in parentheses. Under the settings S1 and S2, the empirical coverage probabilities for Variable 1 are significantly different from the true coverage probabilities (0.9 and 0.95) and the difference is higher in case of high censoring. For scenario S3, the empirical coverage probabilities are close to the true coverage probabilities and this indicates the importance of imputation of the censored observations.}

\begin{table}[ht]
\caption{Average CRPS under different censoring levels L1 and L2 and different settings S1, S2 and S3 based on 100 simulated datasets. The corresponding standard errors are reported in parentheses. A smaller value of average CRPS indicates better performance in spatial prediction.}
\centering
\begin{tabular}{c | ccc}
\hline
\multicolumn{4}{c}{L1: Low-censoring} \\
  \hline
V & S1 & S2 & S3 \\ 
  \hline
1 & 0.593 (0.007) & 0.646 (0.008) & 0.579 (0.006) \\ 
2 & 0.570 (0.006) & 0.591 (0.007) & 0.570 (0.006) \\
   \hline
\multicolumn{4}{c}{L2: High-censoring} \\
  \hline
1 & 0.724 (0.010) & 0.899 (0.013) & 0.591 (0.006) \\ 
2 & 0.571 (0.006) & 0.671 (0.009) & 0.570 (0.006) \\ 
   \hline
\end{tabular}
\label{table2}
\end{table}

\begin{table}[ht]
{\color{black}
\caption{Average coverage probabilities of the 90\% and 95\% prediction intervals under different censoring levels L1 and L2 and different settings S1, S2 and S3 based on 100 simulated datasets. The corresponding standard errors are reported in parentheses.}
\centering
\begin{tabular}{c | ccc | ccc}
  \hline
 & \multicolumn{3}{c}{90\% prediction interval}  & \multicolumn{3}{c}{95\% prediction interval} \\
 \hline
 & \multicolumn{3}{c}{L1: Low-censoring}   & \multicolumn{3}{c}{L1: Low-censoring}  \\
 \hline 
V & S1            & S2            & S3           & S1            & S2            & S3           \\
 \hline
1 & 0.843 (0.006) & 0.809 (0.007) & 0.900 (0.005) & 0.911 (0.005) & 0.885 (0.006) & 0.949 (0.004) \\ 
  2 & 0.903 (0.005) & 0.892 (0.005) & 0.899 (0.005) & 0.954 (0.003) & 0.948 (0.003) & 0.955 (0.003) \\
 \hline
 & \multicolumn{3}{c}{L2: High-censoring}       & \multicolumn{3}{c}{L2: High-censoring}       \\
 \hline
1 & 0.644 (0.008) & 0.612 (0.008) & 0.900 (0.005) & 0.721 (0.007) & 0.701 (0.008) & 0.950 (0.004) \\ 
2 & 0.907 (0.004) & 0.857 (0.007) & 0.901 (0.005) & 0.955 (0.003) & 0.915 (0.005) & 0.955 (0.003) \\
 \hline
\end{tabular}
\label{table3}
}
\end{table}



\section{Data application}
\label{application}

\begin{figure}[h]
\centering
\includegraphics[width = \linewidth]{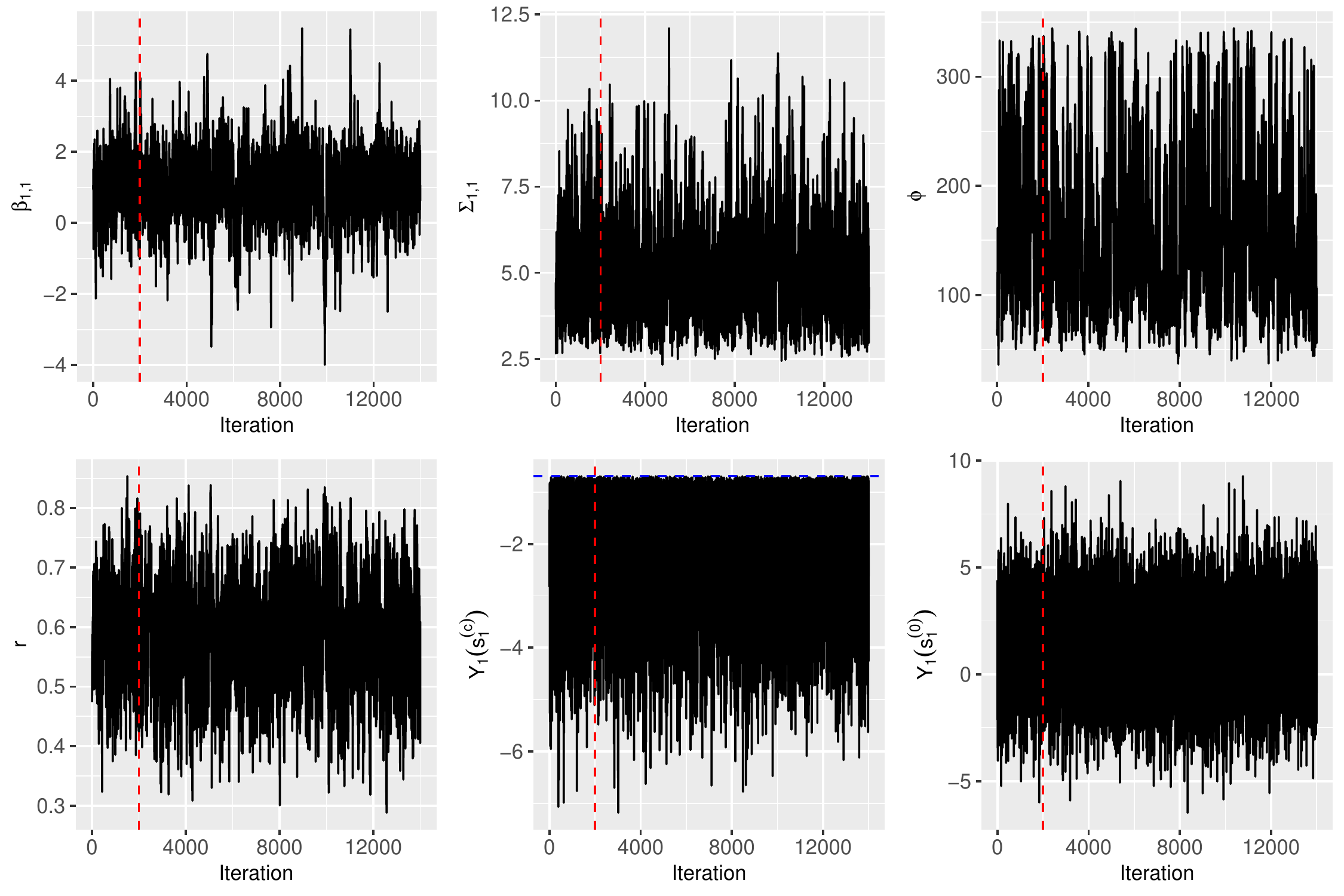}
\caption{Trace plots of some of the model parameters, a censored observation $Y_1(\bm{s}^{(c)}_1)$, and a predicted observation $Y_1(\bm{s}^{(0)}_{1})$. The observations on the left of the red line denote the thinned burn-in samples, and the ones on the right denote the thinned post-burn-in samples. The blue line in the bottom-middle panel indicates the minimum detection limit (on the log scale).}
\label{trace_plots}
\end{figure}

In this section, we illustrate our multivariate Bayesian spatial model by
applying it to the BWDB arsenic contamination dataset described in Section \ref{data}. The trace plots of the MCMC chains presented in Figure \ref{trace_plots} show an overall good mixing and very fast convergence. Additionally, the trace plot of $\phi$ (first row, third column) shows that the estimated range parameter in the model has high variance. The trace plot in the second row, middle column corresponds to a censored observation, on which the minimum detection limit (on the log scale) is shown by the blue line. This trace plot shows that the posterior samples
of $Y_1(\bm{s}_1^{(c)})$ are indeed generated from a truncated posterior distribution. 

\begin{figure}[h]
\centering
\includegraphics[width = \linewidth]{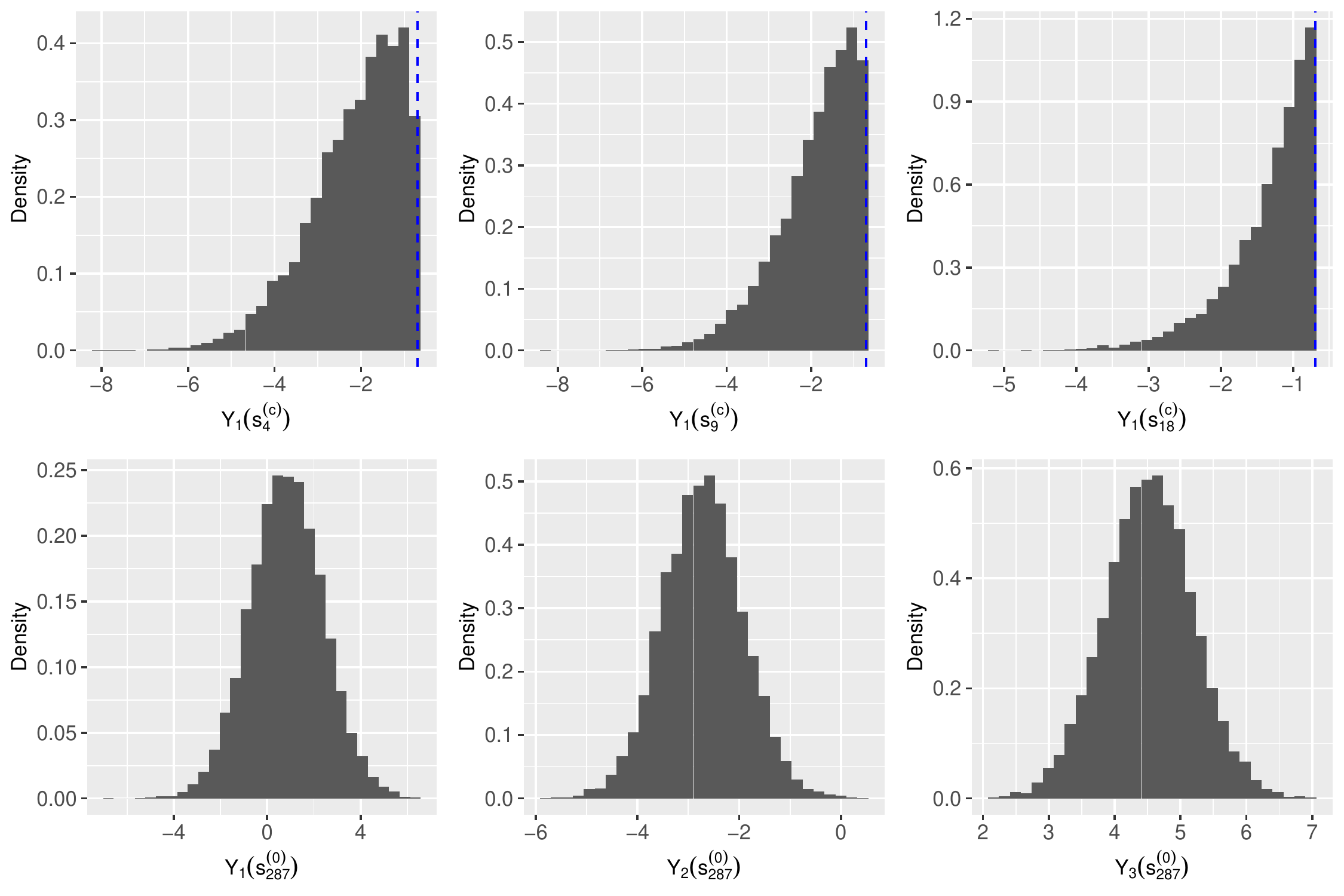}
\caption{First row: Posterior predictive densities of the censored observations at three randomly selected censored locations. The blue lines indicate the minimum detection limit (on the log scale). Second row: Posterior predictive densities of the predicted values at a randomly selected prediction location.}
\label{histograms_censored_predicted}
\end{figure}

Figure \ref{histograms_censored_predicted} (first row) shows the posterior predictive distributions of the censored observations at three randomly selected censored locations. Once again, the blue lines represent the minimum detection limit on the log scale. As expected, the posterior predictive distributions of the censored observations are indeed truncated normal distributions. If the censored observations were replaced by MDL or MDL/2, the problem of estimating these observations would be irrelevant. On the other hand, estimating the censored observations as missing values will give us full posterior predictive distributions thereby ignoring the information that these observations were censored in the first place. Figure \ref{histograms_censored_predicted} (second row) shows the posterior predictive densities of the predicted values for arsenic, barium and calcium concentrations (on the log scale) at a randomly selected prediction location. All the histograms of the posterior predictive samples appear to be unimodal and bell-shaped. 


Table \ref{table4} shows a summary of the posterior inference about the model parameters based on the censored data. The model estimates a positive correlation among the three elements considered. The estimate of the spatial range ($\sim$149 kilometers) suggests a wide spatial dependence among observations. However, the variance associated with this estimate is high. This is quite common in spatial analysis, even with full data, since the likelihood of the range parameter is often quite flat.

\begin{table}[ht] 
\centering
\caption{Posterior means, standard deviations, 0.025-th and 0.975-th quantiles of the model parameters.}
\begin{tabular}{ccccc}
  \hline
Parameter & Mean & SD & 2.5\% & 97.5\% \\ 
  \hline
$\beta_{1,1}$ & 1.00 & 0.95 & -1.08 & 2.82 \\ 
  $\beta_{2,1}$ & -3.10 & 0.44 & -4.08 & -2.29 \\ 
  $\beta_{3,1}$ & 3.12 & 0.40 & 2.19 & 3.84 \\ 
  $\beta_{1,2}$ & 0.29 & 0.59 & -0.87 & 1.45 \\ 
  $\beta_{2,2}$ & -0.44 & 0.30 & -1.03 & 0.17 \\ 
  $\beta_{3,2}$ & -0.75 & 0.25 & -1.25 & -0.27 \\ 
  $\beta_{1,3}$ & -0.24 & 0.57 & -1.36 & 0.91 \\ 
  $\beta_{2,3}$ & -0.41 & 0.29 & -0.98 & 0.15 \\ 
  $\beta_{3,3}$ & -0.48 & 0.24 & -0.95 & 0.00 \\ 
  \hline
  $\Sigma_{1,1}$ & 4.88 & 1.23 & 3.14 & 7.97 \\ 
  $\Sigma_{2,2}$ & 1.25 & 0.30 & 0.82 & 1.99 \\ 
  $\Sigma_{3,3}$ & 0.90 & 0.22 & 0.60 & 1.44 \\ 
  $\Sigma_{1,2}$ & 0.35 & 0.25 & -0.11 & 0.90 \\ 
  $\Sigma_{1,3}$ & 0.16 & 0.21 & -0.24 & 0.59 \\ 
  $\Sigma_{2,3}$ & 0.67 & 0.18 & 0.41 & 1.11 \\
  \hline
  $\phi$ & 148.82 & 66.34 & 59.88 & 306.42 \\ 
  $r$ & 0.59 & 0.09 & 0.41 & 0.75 \\
   \hline
\end{tabular}
\label{table4}
\end{table}

\begin{figure}[!ht]
{\color{black}
\centering
\includegraphics[width = \linewidth]{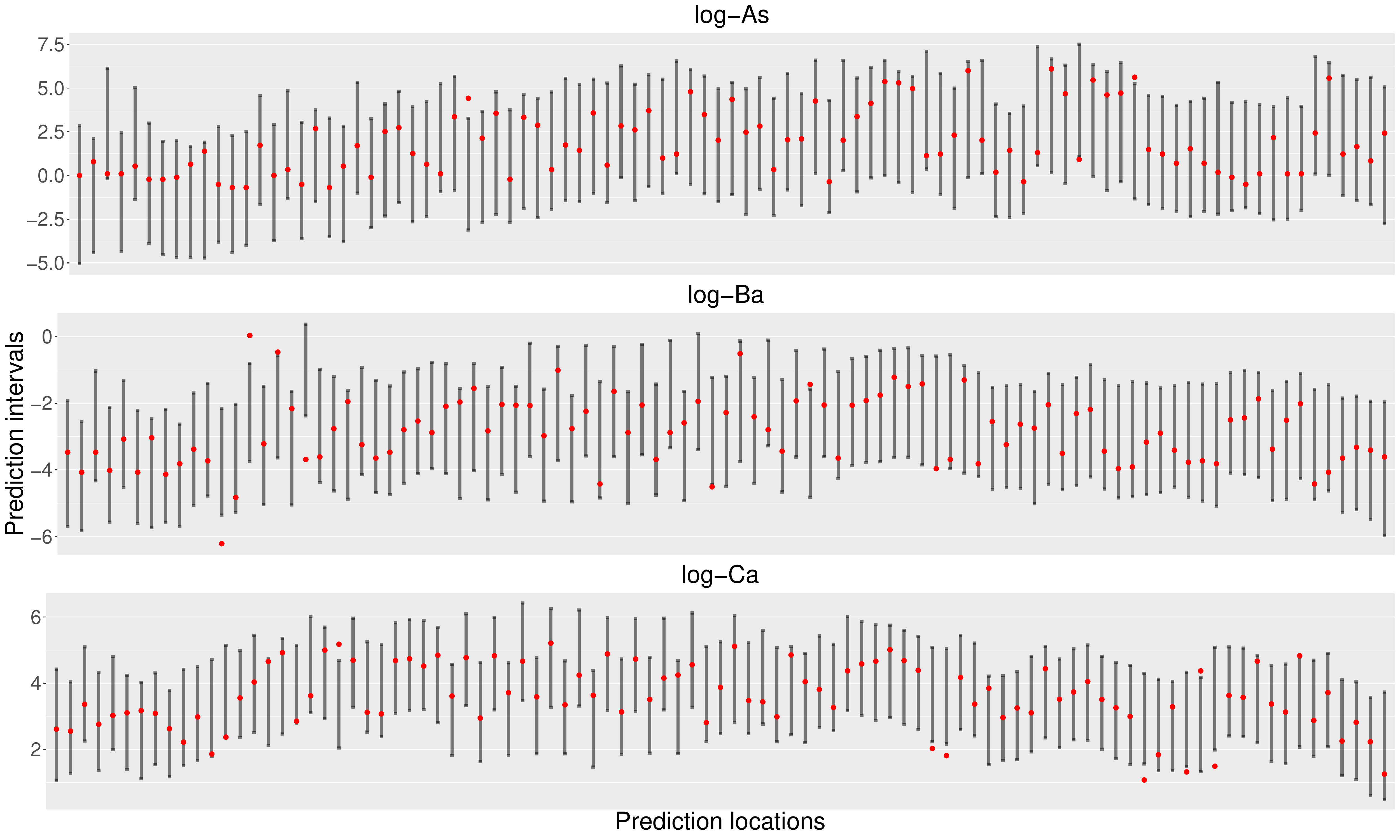}
\caption{Prediction intervals (95\%) for arsenic, barium, and calcium (in log scale), based on leave-one-station-out cross-validation, at 95 sites where the data are fully observed. Red dots indicate the observed values.}
\label{crossvalidation}
}
\end{figure}

\begin{figure}[!ht]
{\color{black}
\centering
\includegraphics[width = \linewidth]{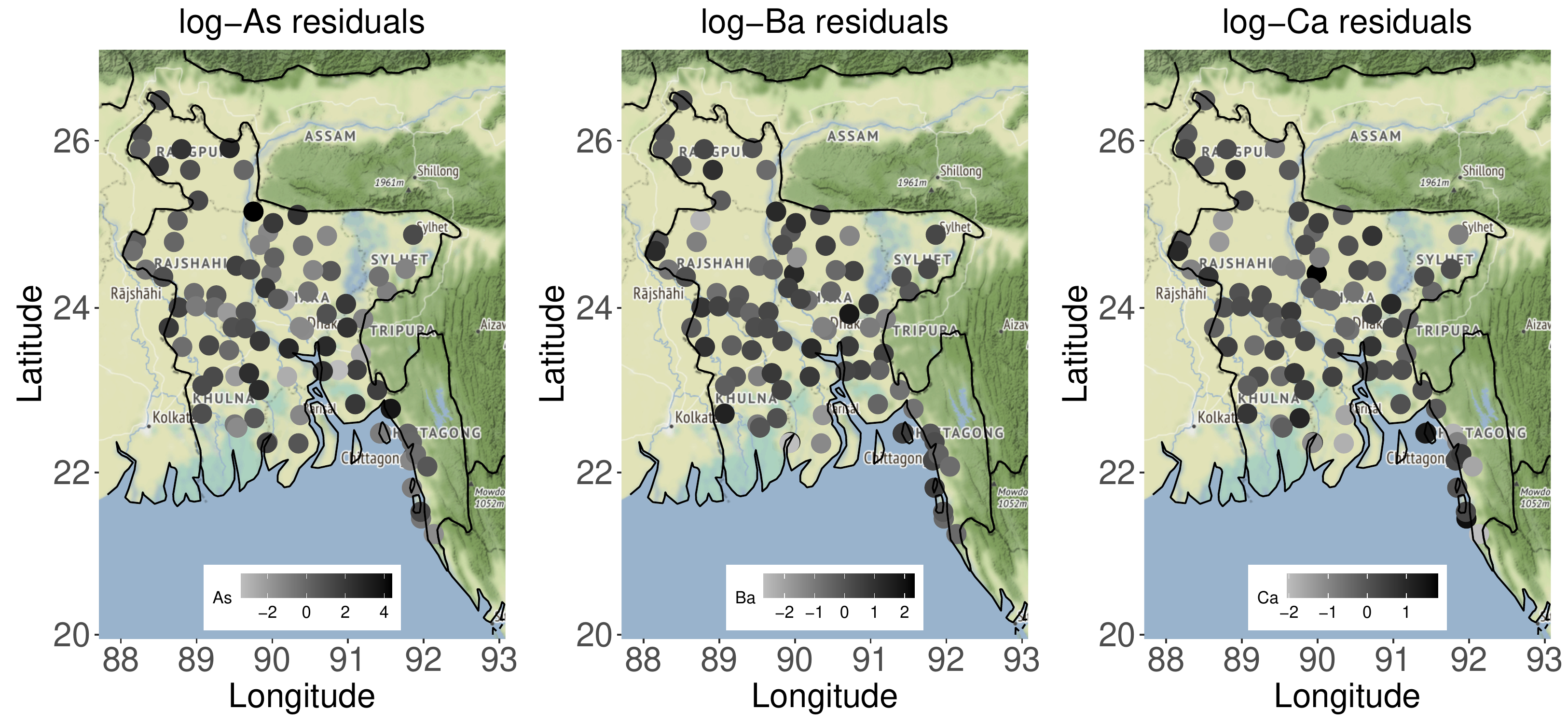}
\caption{Spatial maps of the residuals (observed value minus leave-one-station-out predicted value) of arsenic, barium, and calcium concentrations (in log scale).}
\label{crossvalidation_residuals}
}
\end{figure}

{\color{black} In order to assess the spatial prediction performance of the proposed statistical model, we perform a leave-one-station-out cross-validation. We consider the 95 spatial locations where the data are fully observed and remove one of them at a time to re-fit the model to the rest of the data (including the censored observations). For each cross-validation iteration, we obtain 95\% prediction intervals of arsenic, barium, and calcium concentrations (in log scale) at the removed site. The prediction intervals and the observed values at the test sites are presented in Figure \ref{crossvalidation}. For most of the cases, the prediction intervals include the observed values. This indicates that the model (\ref{main_model}) performs well in terms of spatial prediction. 

The spatial maps of the residuals (the observed value minus leave-one-station-out predicted value) are presented in Figure \ref{crossvalidation_residuals}. The residuals vary across a large range of values and no clear spatial trend is observed for any of the variables. For example, the residual at $89.751^\circ$E and $25.156^\circ$N is highly positive (4.38). Figure \ref{spatial_maps} shows that the arsenic concentration level at the nearby stations are substantially small, with most of them being below the MDL (0.5 $\mu$g). However, the arsenic concentration at that site is 82.50 $\mu$g, which is $\exp[4.38] \approx 80$ times higher than the predicted value. The model (\ref{main_model}) fails to capture such high nonstationarity; however, nonstationary spatial models can lead to spurious estimates when the inference is drawn based on only a limited number of observations (in our case, for example). Incorporating important covariates can be a solution in this context; this specific site is located near the confluence of the rivers Teesta and Bramhaputra, and thus, soil features could possibly explain the high variability in mineral concentration of groundwater.}

\begin{figure}[!ht]
\centering
\includegraphics[height = 1.2\linewidth]{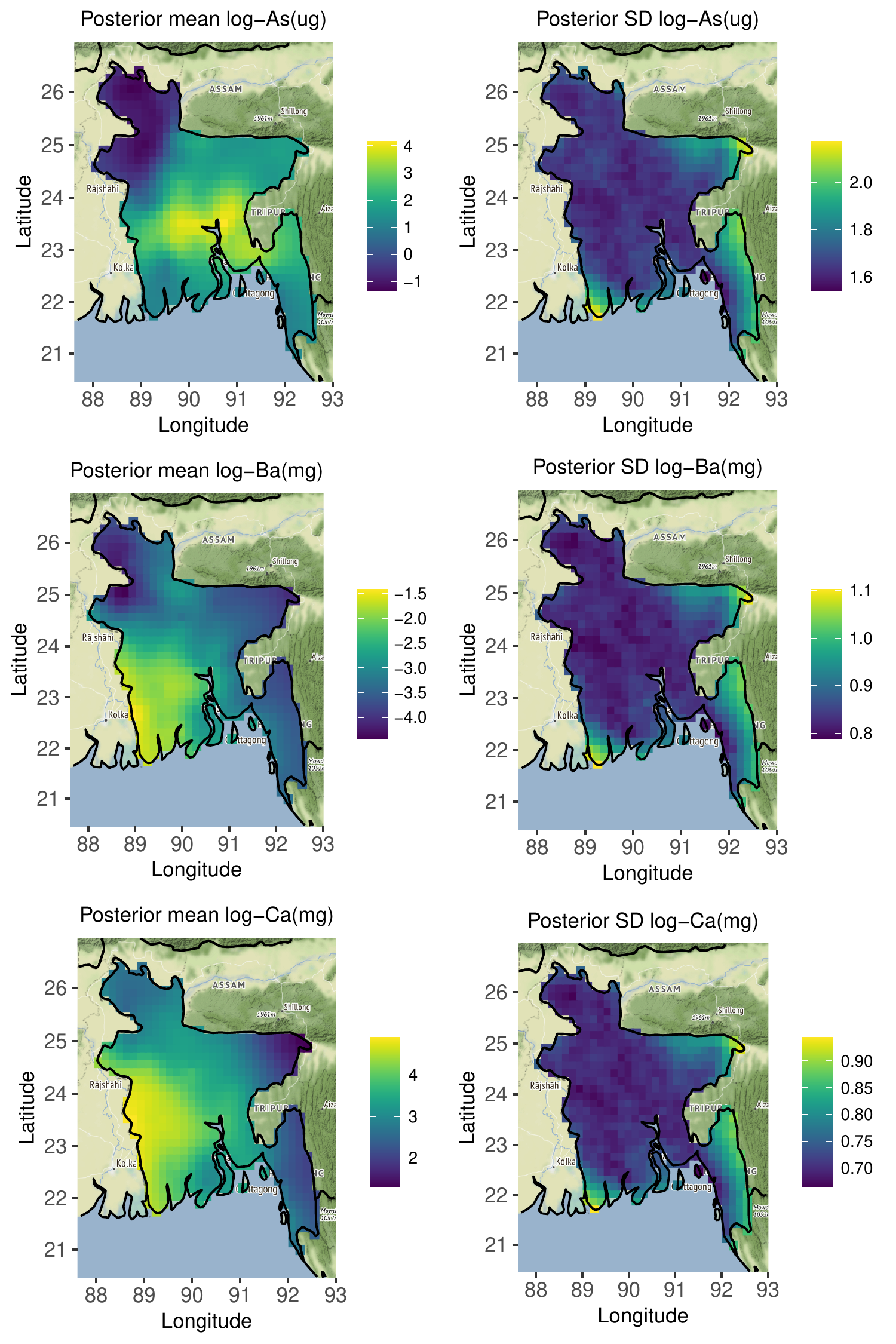}
\caption{First column: Prediction maps of arsenic, barium and calcium concentrations (on the log scale) over Bangladesh using posterior means of the respective prediction distributions. Second column: Uncertainties associated with the prediction calculated using posterior standard deviations (SD) of the respective prediction distributions. }
\label{spatial_maps_predicted}
\end{figure}

Figure \ref{spatial_maps_predicted} (first column) shows the spatial maps for arsenic, barium, and calcium (on the log scale) over Bangladesh calculated using the mean of the posterior predictive distributions. The second column of Figure \ref{spatial_maps_predicted} shows the associated uncertainties in prediction calculated using the standard deviations of the posterior predictive samples. Based on these maps, high levels of arsenic contamination are seen in the divisions of Dhaka, Khulna, and the northwestern part of Chittagong, whereas moderate arsenic contamination is seen in parts of Sylhet and north-eastern Chittagong. Only the division of Rangpur and parts of Rajshahi in the north-western part of Bangladesh register a low concentration of arsenic. The spatial maps also highlight the positive correlation among concentrations of arsenic, barium, and calcium. Not surprisingly, the uncertainties associated with the predictions are low in areas where observations are present, whereas the uncertainties are higher in regions with no observations.

We also draw inferences about division-wise mean contamination levels for the seven divisions of Bangladesh. First, we discretize the spatial domain into a grid of 526 prediction locations as considered in Figure \ref{spatial_maps_predicted}. Further, we divide them into seven regions according to the divisional boundaries obtained from \url{https://rpubs.com/asrafur_ashiq/map_of_bangladesh}. We denote the spatial domain of $j$-th division by $\mathcal{S}_j$. The mean contamination level of the $p$-th element within $\mathcal{S}_j$ is $M_{jp} = \vert \mathcal{S}_j \vert^{-1} \int_{\mathcal{S}_j} Y_p(\bm{s}) d \bm{s} $. This integral is approximated by $M_{jp} \approx N_j^{-1} \sum_{\bm{s}^{(0)}_m \in \mathcal{S}_j} Y_p(\bm{s})$, where $N_j$ denotes the number of prediction locations within $\mathcal{S}_j$. We calculate the posterior means and the corresponding standard errors from the posterior predictive samples of $\{Y_p(\bm{s}), \bm{s} \in \mathcal{S}^{(0)}\}$ and report them in Table \ref{divisionwise_means}. 

\begin{table}[ht]
\centering
\caption{Division-wise posterior mean contamination levels and the corresponding standard errors.}
\begin{tabular}{llll}
  \hline
 & As($\mu$g/L) & Ba(mg/L) & Ca(mg/L) \\ 
  \hline
  Barisal & 31.864 (0.35) & 0.121 (0.0004) & 38.326 (0.11) \\ 
  Chittagong & 68.201 (0.52) & 0.056 (0.0001) & 25.499 (0.05) \\ 
  Dhaka & 65.669 (0.40) & 0.099 (0.0002) & 55.385 (0.08) \\ 
  Khulna & 31.385 (0.37) & 0.239 (0.0007) & 118.923 (0.25) \\ 
  Rajshahi & 8.306 (0.07) & 0.060 (0.0001) & 68.102 (0.12) \\ 
  Rangpur & 1.856 (0.01) & 0.038 (0.0001) & 22.024 (0.04) \\ 
  Sylhet & 31.050 (0.59) & 0.038 (0.0002) & 12.359 (0.04) \\ 
   \hline
\end{tabular}
\label{divisionwise_means}
\end{table}

These results corroborate with the spatial maps seen in Figure \ref{spatial_maps_predicted}. The divisions of Chittagong and Dhaka have the highest mean concentrations of arsenic in the groundwater followed by Barisal, Khulna and Sylhet. Only Rajshahi and Rangpur have concentrations of arsenic which are below the current permissible limit of arsenic in drinking water (10 $\mu$g/L), as prescribed by the World Health Organization (WHO). This shows that more than 76\% of the total population in Bangladesh (approximately 110 million people) is exposed to toxic levels of arsenic concentration in their drinking water. The Bangladesh population estimates were obtained from the Population Monograph of Bangladesh published by the Bangladesh Bureau of Statistics (BBS) in November 2015.

\section{Discussions and conclusions}
\label{discussion}

The arsenic contamination in Bangladesh is potentially the largest naturally occurring environmental disaster in human history. The complex spatial pattern of arsenic abundance and its relationship with other contaminants makes the problem even more severe. To this end, this work presents a multivariate spatial Bayesian framework for joint modeling of the concentrations of contaminants in groundwater in the presence of left-censored observations.
Inference about model parameters, including all censored data, is based on an adaptive MCMC. The nugget effect present in the proposed model naturally handles all censored observations and allows univariate updates, thereby avoiding any computational burden associated with multivariate likelihoods for censored observations. Computer Codes (written in R) used in this paper are available at \url{https://github.com/arnabstatswithR/Arsenic-contamination-mapping.git}.

Several extensions can be made to the proposed model to add more flexibility to the model structure. Here, the covariance function is assumed to be separable; however, the model can be extended to incorporate non-separable covariance models. Also, to keep notations simple, we have assumed that $Y_1(\cdot)$ is left-censored at a censoring level $u$. Generally, while considering contamination data, the level of censoring depends on the site from which the data has been collected, due to varying precision levels of the data collecting instruments at different locations. Extending the proposed model to incorporate site-dependent minimum detection limits is straightforward, where instead of drawing posterior samples from the truncated normal distribution with common truncation limit $u$, the samples will be drawn from truncated normal distributions with truncation limits $u(\bm{s}^{c})$, where $\bm{s}^{c}$ denotes a location with censored observation. The model can also be easily modified to incorporate right-censored or interval-censored data.

{\color{black} The inclusion of covariates in the spatial model for arsenic concentration depends on the overall goal of the study. In our study, the goal is to make spatial maps of arsenic concentration over Bangladesh. In this case, the only covariates used were latitudes and longitudes, since data on other covariates were not available at the prediction locations. However, if the goal of the study is to quantify the effect of covariates on the spatial distribution, several covariate information can be used. These include hydrogeological variables such as well depth, hydrodynamic variables such as mean groundwater fluctuation and geographical and seasonal variables such as latitude, longitude, elevation and seasonality. See \cite{shamsudduha2015generalized} for a full discussion on the rationale for considering different covariates while analyzing arsenic variations in the groundwater of Bangladesh.}

A large contamination dataset resulting from a systematic survey of 61 of the 64 districts of Bangladesh conducted by the British Geological Survey, involving a collection of groundwater samples from 3534 boreholes is also available on the same website as mentioned in Section \ref{data}. However, the proposed hierarchical Bayesian framework is not scalable to densely collected data. The issue of handling large datasets has been studied extensively in spatial statistics and more sophisticated methods incorporating fast approximation algorithms can be developed for such datasets. Finally, future work could also focus on data fusion, that is, merging datasets from different sources and modeling arsenic abundance based on the combined dataset.

\section*{Acknowledgement}

The authors would like to thank the Special Issue Editor Snigdhansu Chatterjee from University of Minnesota, United States, and two anonymous reviewers for their suggestions. The second author would also like to thank Rapha{\"e}l Huser from KAUST, Saudi Arabia.









\bibliographystyle{apalike}
\bibliography{jisaref}







\newpage

\noindent 
{\bf Indranil Sahoo} \\
Room 4127, Grace E. Harris Hall\\
Department of Statistical Sciences and Operations Research\\
Virginia Commonwealth University \\
Richmond, United States 23284. \\
E-mail: sahooi@vcu.edu\\

\vspace{.1in}
  
\noindent 
{\bf Arnab Hazra}\\  
Room No. 4200-CU07, Building 1\\
Computer, Electrical and Mathematical Sciences and Engineering Division \\
King Abdullah University of Science and Technology \\
Thuwal, Saudi Arabia 23955.\\
E-mail: arnab.hazra@kaust.edu.sa

\end{document}